\documentclass[twocolumn,showpacs,preprintnumbers,amsmath,amssymb]{revtex4}
\usepackage{graphicx}
\usepackage{dcolumn}
\usepackage{bm}

\begin{document}

\preprint{APS/123-QED}

\title{The effective $S$-matrix from conductance data\\ 
in a quantum wave guide: Distinguishing the indistinguishable}

\author{Gursoy B. Akguc$^1$}
\author{Jorge Flores$^1$}
\author{Sergey Yu. Kun$^{1,2,3}$}
\affiliation{$^1$Centro de Ciencias F\'{i}sicas, Universidad
Nacional Aut\'{o}noma de  M\'{e}xico, \\
Cuernavaca, Morelos, M\'{e}xico}
\affiliation{$^2$Center for Nonlinear Physics, RSPhysSE,\\
The Australian National University, Canberra ACT 0200, Australia}
\affiliation{$^3$Department of Theoretical Physics, RSPhysSE\\
The Australian National University, Canberra ACT 0200, Australia}

\date{\today}

\begin{abstract}
We consider two different stationary random processes  whose probability
distributions are very close and indistinguishable by standard tests 
for large but
limited statistics. Yet we demonstrate that these processes can be
reliably distinguished. The method is applied to analyze conductance 
fluctuations in coherent electron transport through nanostructures.
\end{abstract}

\maketitle

The quantum-mechanical scattering by complex systems
has been a problem of long-standing interest. This problem
is of a great importance in nuclear, molecular and mesoscopic physics
\cite{kn:gmg,kn:been,kn:alhassid,kn:bird}.
The scattering process can be often analyzed in terms 
of two distinct time scales: (i) a prompt response due to direct
processes and (ii) a time-delayed scattering mechanism. In nuclear
physics the time-delayed mechanism is associated with the formation of an 
equilibrated compound nucleus \cite{kn:gmg, kn:eric}. Similarly,
 time-delayed processes have been encountered in the study of coherent 
electron transport through nanostructures \cite{kn:gmg,kn:been,kn:alhassid,
kn:bird, kn:akg1}. The two distinct
time scales introduce two different energy scales. Namely, the 
scattering amplitude of direct processes is a smooth function of energy
and it can be taken as an energy averaged $S$-matrix. The scattering amplitude
of time-delayed process, on the other hand, is an energy dependent function 
which fluctuates rapidly 
around zero. Knowledge of the energy averaged $S$-matrix 
allows one to construct a probability distribution (PD) of the 
full $S$-matrix \cite{kn:mello, kn:gopar, kn:akg1}.

$S$-matrix for ballistic electron scattering off nanostructures
can be calculated numerically if the geometry of the nanodevice, as well as
the shape,
size and number of propagating modes of the leads, and spatial distribution 
of disorder are known precisely.
Then, having calculated the energy dependence of the $S$-matrix one can
directly separate it into an energy averaged direct reaction component
and a fluctuating one. In experiments, the detailed information 
about the conducting device may not be always available. For example, 
wall imperfections and unknown spacial distribution of disorder
 does not allow to evaluate the contribution of direct processes.
 Disorder can also block leads in an unknown manner reducing the number of 
propagating modes. This motivates us to consider the ``inverse problem'': 
What information about relative contributions of direct and time-delayed
processes and number of propagating modes in the leads 
can be obtained from the analysis of the transmission
energy fluctuations provided the above
 mentioned characteristics of the nanodevice are unknown? 

For concreteness we consider the transmission between horizontally oriented
leads for the two types of geometry of wave guides (WG) (Fig. 1).
For experimental purposes,
we refer the reader to reference  \cite{kn:akg2} for a possible 
implementation of this system.
The left
leads of both WG can accommodate a single propagating channel only. 
 For the WG (b) the right lead also accommodates only one channel 
($N=1$), while
for WG (a) it has $N\geq 2$ open channels. In addition, both WG (a) and (b)
have $N_a\gg 1$ and $N_b\gg 1$ open channels shown by arrows pointing down 
at the bottom of WG with $N_a\gg N$. 
Disorder  
is modeled by scatterers (shown in Fig. 1 in white) having 
infinitely high potential walls.

\begin{figure}
{\centering \scalebox{0.5}{\includegraphics{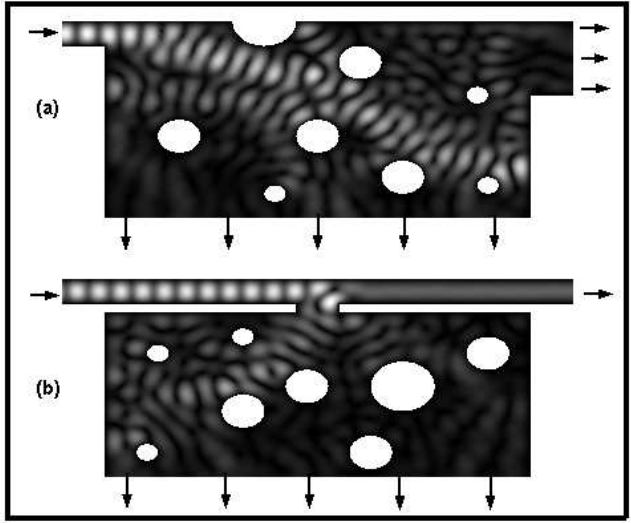}}\par }
\caption{\label{fig:f1} The electron probability distribution inside the 
wave guides for an arbitrary chosen electron energy.
The arrows show incident unit current from left and current leaving 
the system from the right. The white circular regions represent infinite 
potential barriers.
The wave guide (a) does not show direct processes and (b) has direct 
processes (see text).}
\end{figure}

Transmission between the horizontally oriented leads for the (a) and (b)
WG in Fig. 1 is given by
\begin{equation}
T(E)=\sum_{j=1}^N|S_{1j}(E)|^2.
\end{equation}
Here $S_{1j}(E)$ are the $S$-matrix elements for scattering between
left leads ``1'' and right leads $j$, and $E$ is electron energy. 
For the WG (b) in Fig. 1
the sum (1) contain only a single term ($N=1$).
We use the decomposition $S_{1j}(E)=<S_{1j}>+\delta S_{1j}(E)$, where
$<S_{1j}>$ are energy averaged $S$-matrix elements, and
$\delta S_{1j}(E)$ are energy fluctuating ones. We chose the spacial
distribution of disorder such that (i) $\sum_{j}|<S_{1j}>|^2\ll 1$, and (ii)
transmission between the horizontal leads and open channels at the 
bottom of both (a) and (b) WG in Fig. 1 is mostly due to time-delayed
processes. Therefore, the overall amount of direct processes for the 
transmission between different channels is much less than
that for the overall amount of time-delayed processes.
Also, due to the presence of disorder, the dynamics of the classical 
counterpart
in the interaction region of WG (a) and (b) in Fig. 1 is chaotic. Since
the total numbers of open channels, $N+N_a+1$ and $N+2$, are much greater
than unity, we are in the regime of Ericson fluctuations when 
electron resonance states within the WG are overlapping.  
 Then each individual fluctuating $S$-matrix element
can be considered as a Gaussian random process.
Namely, real and imaginary
parts of each individual $\delta S_{1j}(E)$ in sum (1) are distributed
by a Gaussian law, are uncorrelated and have the same dispersions 
\cite{kn:gmg,kn:eric}.

The PD for
 $y=T(E)/<T(E)>$ (so that $<y>=1$)
is given by \cite{kn:eric},
\begin{eqnarray}
P_N(y)&=& N(1-y_d)^{-1}(y/y_d)^{(N-1)/2}\nonumber\\
&&\exp[-N(y_d+y)/(1-y_d)] \nonumber \\
&&I_{N-1}[2N(y_dy)^{1/2}/(1-y_d)].
\end{eqnarray}   
Here $I_{N-1}(\lambda )$ is the modified Bessel function of 
$(N-1)$ order and 
$y_d=\sum_{j}|<S_{1j}>|^2/<T(E)>$ is the relative contribution of direct
processes to the total transmission. 
The PD of Eq. (2) is exact 
if all $\delta S_{1j}(E)$ with different $j$ are uncorrelated
and have the same dispersions. Otherwise this PD is an approximation
for which $N$ should be understood as an effective number of
independent channels $N_{eff}< N$. It is given by
the normalized variance of 
 $\delta T(E)=\sum_{j=1}^N|\delta S_{1j}(E)|^2$:
 $<\delta T(E)^2>/<\delta T(E)>^2
-1=1/N_{eff}$.

For the WG (b) in Fig. 1 one has to put
$N=1$ in Eq. (2). For $y_d=0$, Eq. (2) takes the form
of a $\chi^2$-distribution with
$2N$ degrees of freedom \cite{kn:eric}:
$$
P_N(y)=N[\Gamma (N)]^{-1}(Ny)^{N-1}\exp(-Ny),
\eqno{(3)}
$$
where $\Gamma (N)$ is Gamma function.  
Note that the variance of $y$ is given by  
$<y^2>-1=(1-y_d^2)/N$ \cite{kn:eric}.

As an example, we consider two possible cases for WG (a) in Fig. 1: 
(i) $N=10$ and $y_d=0$, the latter due 
to complete blocking of the direct
paths between the leads by disorder, and (ii) $N=5$ and 
$y_d=(0.8)^{1/2}$. Also
 suppose that, for WG (b), for which
$N=1$, $y_d=(0.9)^{1/2}$. For the above three sets
of $N$ and $y_d$ values the variance of conductance $y$ is the same and 
equals
0.1. The question is: Can we distinguish these three cases by analyzing
PD of $y$ in Eq. (2)?

\begin{figure}
{\centering \scalebox{0.5}{\includegraphics{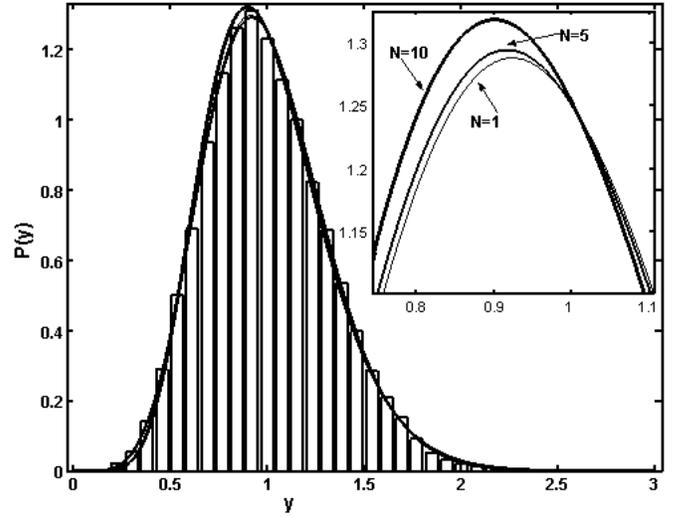}}\par }
\caption{\label{fig:f2} The probability distributions of transmission.
The solid lines, with decreasing thicknesses, show 
the theoretical curves for $N=1$, $N=5$ (Eq. (2)) and $N=10$ (Eq. (3)) 
channel cases respectively. The histogram is constructed from the numerical
data for $N=1$ and $y_d=(0.9)^{1/2}$, as an example.
 The inset shows a blow up of the transmission probability
distributions demonstrating that the difference between them 
around their maxima does not exceed 4$\%$.}
\end{figure}

In Fig.~\ref{fig:f2} we plot the PD for the three sets of $N$ and $y_d$.
The three distributions are very close.
We performed a $\chi^2$ test and found that, even for as many as 28000 
independent realizations of $y$
for each of the three
cases, the distributions are indistinguishable at a $99\%$ confidence level.
And, to the best of our knowledge, there is currently 
 no any other statistical test which would allow to distinguish 
between the   
three different stochastic processes.
Yet, in what follows, we show that the problem of distinguishing between
these random processes can be solved.

We will refer to the $y$ as  $y_i$, where
the index $i$ stands for different independent realizations of the process.
Let us transform to new random variables
$s_{ij}=(y_i+y_j)/2$, $r_{ij}=(y_i-y_j)/2s_{ij}$ with $i\neq j$. 
One can now easily find the joint probability distribution $K(s,r)$.
The analytical calculations are formally similar to those of Ref.~\cite{kn:kun}
for the analysis of the correlation between cross section and analyzing
power in the regime of Ericson fluctuations in nuclear reactions.
Without presenting the explicit result for $K(s,r)$ we point out that,
for $y_d=0$, $s$ and $r$ are statistically independent. 
On the contrary, for $y_d\neq 0$, $s$ and $r$ correlate, the bigger
$y_d$ the stronger this correlation is.

\begin{figure}
{\centering \scalebox{0.5}{\includegraphics{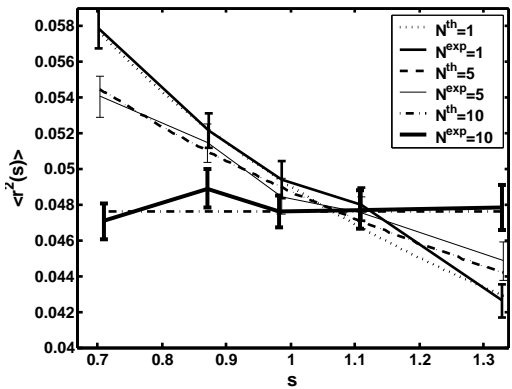}}\par }
\caption{\label{fig:f3} The plot of $<r^2(s)>$ vs. $s$ corresponding to 
the same $N$ (shown in inset) and $y_d$ values as those used for the 
analysis of the probability distributions in Fig.~\ref{fig:f2}.
Dotted, dashed and dotted-dashed lines are theoretical predictions 
(Eq. (4).}
\end{figure}

We calculate  $<r^2(s)>$:
$$
<r^2(s)>=[1-I_{2N+3}(\lambda )/I_{2N-1}(\lambda )]/(2N+1),
\eqno{(4)}
$$
where $\lambda= 4N(y_d s)^{1/2}/(1-y_d)$, and
 take into account the constraint $(1-y_d^2)/N=0.1$.
In Fig.~\ref{fig:f3}, we show $<r^2(s)>$ for the three cases. 
We have generated 28000 realizations for each case and used
all possible $i< j$ combinations for $s_{ij}$ and $r_{ij}$. 
The data in
Fig.~\ref{fig:f3} are obtained by dividing the $s_{ij}$ into 5 bins 
for each case and taking the average of $r^2(s)$ and $s$ for each bin.
It is evident from Fig.~\ref{fig:f3} that we can distinguish the three 
cases. 
The 
fluctuations around the theoretical curves are determined by the number of 
realizations. 
Error bars are calculated numerically from
standard
 deviations of 100 data sets each consisting 28000 $y$ realizations.
 We found
that it is possible to distinguish the $N=1$ and $N=10$ processes in $97\%$
of cases for 210 realizations of $y$. 
This drops to $85\%$ for 140 realizations.

We apply the new method to distinguish between (a) and (b) in Fig. 1
from the conductance fluctuation data. 
We implemented a finite element solution of the Schr\"odinger equation 
\cite{kn:akg2, kn:akg3}.
\begin{figure}
{\centering \scalebox{0.4}{\includegraphics{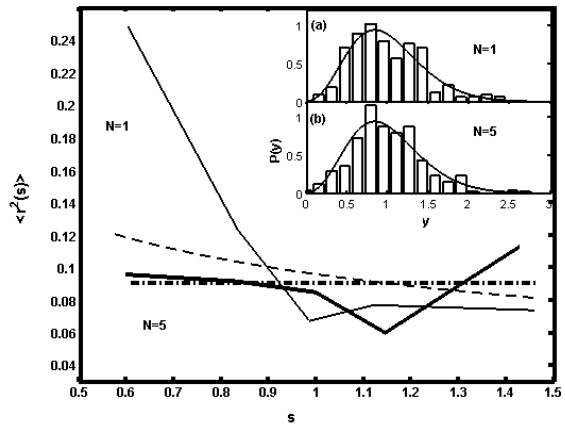}}\par }
\caption{\label{fig:f4} The plot of $<r^2(s)>$ vs. $s$ for the 
wave guides shown in Fig.~\ref{fig:f1}. The dashed and dotted-dashed lines
are theoretical estimates (Eq. (4)) and thin and thick solid lines 
are numerical
calculations for $N=1$ (with direct processes) and for $N=5$ (without direct
processes) cases respectively.
The inset shows conductance distributions (lines) 
for (a) $N=1$, with direct processes (Eq. (2)),
and (b) $N=5$, without direct processes (Eq. (3)).
Histograms are constructed from the numerical data.}
\end{figure}
Calculations were performed on the energy range which allows 5 propagating
channels in the right lead for WG (a) in Fig. 1. Both left leads for 
WG (a) and (b) as well as the right lead for WG (b) in Fig. 1 accommodate 
a single channel. 
In addition, for both (a) and (b) WG in Fig. 1, there are 20-25 propagating 
channels from the lower 
part of the devices.
A unit current is directed from left. 
We are interested in the transmission 
probability from the left to the right side. 
 Spacial disorder distribution
for WG (a) in Fig. 1 was chosen to block the direct paths between the
left and right leads, i.e. to maximally suppress direct processes.
This was confirmed by $S$-matrix numerical data, which yielded
negligible absolute value of the energy averaged $S$-matrix, and by the fact
that the normalized variance of conductance fluctuations is close to 0.2.
For WG (b) in Fig. 1 the extension of leads into the cavity is
adjusted to have relative contribution of direct process $y_d\approx 0.9$ 
so that the normalized variance of conductance fluctuations is close to 0.2,
i.e. to that for WG (a) in Fig. 1. For both (a) and (b) WG in Fig. 1
we used different disorder distributions to increase the statistics.
The overall number of independent realizations of transmission values $y$
for each type of geometry in Fig. 1 was around 150.

In Fig. 4 we present PD of $y$ for the (a) and (b) WG of Fig. 1. 
$\chi ^2$ test does
not allow to distinguish the two PD. Yet, from Fig. 4 one can see
that for the $N=1$ case, in the presence of direct processes, $<r^2(s)>$ 
strongly increases when $s$ decreases. On the contrary, for the $N=5$
case, when direct processes are absent, $<r^2(s)>$ for $s<1$ is close
to a constant. We have checked that even for each single configuration
of disorder, when we have only about 30-40 independent
realizations of the transmission values for each (a) and (b) WG,
$<r^2(s)>$ increases always noticeably faster for the $N=1$ case with
direct processes as compared to the $N=5$ case without direct processes.
On the other hand, the rise of $<r^2(s)>$ when $s$ decreases of $s$ 
for $N=1$ with direct processes
 is much stronger than that predicted by Eq. (4).
In order to find a possible reason for this we analyzed the 
PD of fluctuating $S$-matrix elements. We found that (i) $\delta S_{11}(E)$
 are not distributed isotropically in the complex plane, (ii)
the PD of real and imaginary parts of  $\delta S_{11}(E)$ 
are not Gaussian, and (iii) real and imaginary parts of  $\delta S_{11}(E)$
are correlated. Therefore, the reason for the discrepancy between
the data and theoretical line in Fig. 4 for $N=1$ is that the conditions to 
derive Eq. (4) are not met. Yet, the distribution for transmission is not 
sensitive
to the deviation of the PD of  $\delta S_{11}(E)$ from its isotropic Gaussian
distribution in the complex plane and correlation between real and imaginary
parts of $\delta S_{11}(E)$.

In conclusion, we have suggested a method to distinguish between
 different stationary random processes  whose PD
 are very close and indistinguishable by standard tests 
for large but limited statistics.  
 The method has been applied to analyze conductance 
fluctuations in coherent electron transport through WG.
We found that the method proposed here, unlike the analysis of PD
of transmission, is sensitive to (i) the deviation of PD of
fluctuating transmission amplitudes from isotropic Gaussian statistics 
in the complex plane, and (ii) correlations between 
real and imaginary parts of the fluctuating transmission amplitudes.

\end{document}